\documentclass[journal,comsoc]{IEEEtran}
\usepackage[T1]{fontenc}
\usepackage{graphicx}
\usepackage{comment}
\usepackage{mathtools}
\usepackage{multirow}
\usepackage{tikz}
\newcommand*\circled[1]{\tikz[baseline=(char.base)]{
            \node[shape=circle,draw,inner sep=.5pt] (char) {#1};}}
\usepackage[pagebackref=true,breaklinks=true,colorlinks,bookmarks=false]{hyperref}
\usepackage{color, colortbl}
\definecolor{Gray}{gray}{0.9}
\definecolor{LightCyan}{rgb}{0.88,1,1}
\ifCLASSINFOpdf
\else
 
\fi
\usepackage{amsmath}
\usepackage[cmintegrals]{newtxmath}
\begin{document}
\title{SRTransGAN: Image Super-Resolution using Transformer based Generative Adversarial Network}

\author{Neeraj Baghel, Shiv Ram Dubey, Satish Kumar Singh
\thanks{N. Baghel, S.R. Dubey and S.K. Singh are with the Computer Vision and Biometrics Lab at Department of Information Technology, Indian Institute of Information Technology, Allahabad-211015, India (email: neerajbaghel@ieee.org, srdubey@iiita.ac.in, sk.singh@iiita.ac.in).
}%
}

\markboth{SRTransGAN}%
{Baghel \MakeLowercase{\textit{et al.}}}

\maketitle

\begin{abstract}
Image super-resolution aims to synthesize high-resolution image from a low-resolution image. It is an active area to overcome the resolution limitations in several applications like low-resolution object-recognition, medical image enhancement, etc. The generative adversarial network (GAN) based methods have been the state-of-the-art for image super-resolution by utilizing the convolutional neural networks (CNNs) based generator and discriminator networks. However, the CNNs are not able to exploit the global information very effectively in contrast to the transformers, which are the recent breakthrough in deep learning by exploiting the self-attention mechanism. Motivated from the success of transformers in language and vision applications, we propose a SRTransGAN for image super-resolution using transformer based GAN. Specifically, we propose a novel transformer-based encoder-decoder network as a generator to generate $2\times$ images and $4\times$ images. We design the discriminator network using vision transformer which uses the image as sequence of patches and hence useful for binary classification between synthesized and real high-resolution images. 
The proposed SRTransGAN outperforms the existing methods by 4.38\% on an average of PSNR and SSIM scores. We also analyze the saliency map to understand the learning ability of the proposed method. 
\end{abstract}


\IEEEpeerreviewmaketitle

\section{Introduction}
\IEEEPARstart{I}{mages} are the most common data used in the digital world. Millions of digital images are being captured by cameras, CCTVs, and mobiles in high-resolution. These images are used for various applications for recognising persons, objects, textures, scenes, etc. However, the storage, transmission and processing of high-resolution images at scale pose a great challenge in terms of space, bandwidth and computational requirements.
To solve these problems high-resolution images are generally down-sampled in low-resolution for storage, transmission and processing purposes. Many times the appearance of the concerned object is low-dimensional even in high-dimension images, such as surveillance images. The performance of the computer vision methods suffer due to the lack of sufficient information in the low-resolution images. Hence, a common practice is to increase the resolution of images using up-sampling when required, but it is a lossy process. For problems like recognising far objects in low-resolution, the up-sampling and image enhancement not provide the satisfactory results \cite{ed_chen2021multispectral, ed_huang2021effective}.
Recently, image super-resolution has got a great attention \cite{survey2020IeeeTransPatten, ieee_imsur}. The image super-resolution has shown a tremendous growth using deep learning approaches, such as convolutional neural networks (CNNs) \cite{SRCNNr1, FSRCNNr2, VDSRr3, Memnet}, generative adversarial networks (GANs) \cite{SRGAN,esrgan,GMGAN}, recently developed transformer networks \cite{ESRT, RCAN, IPT}. The image super-resolution approaches for specific applications have also been developed, such as face, spatiospectral \& thermal image super-
resolutionface super-resolution \& medical image super-resolution \cite{GCFSR, SpatiospectralSR, ThermalSR}. These methods learn  the mapping function between the low and high resolution images which helps to generate a super-resolution image for the low resolution counterpart. 


\begin{figure}[t]
\centering
\includegraphics[width=\columnwidth]{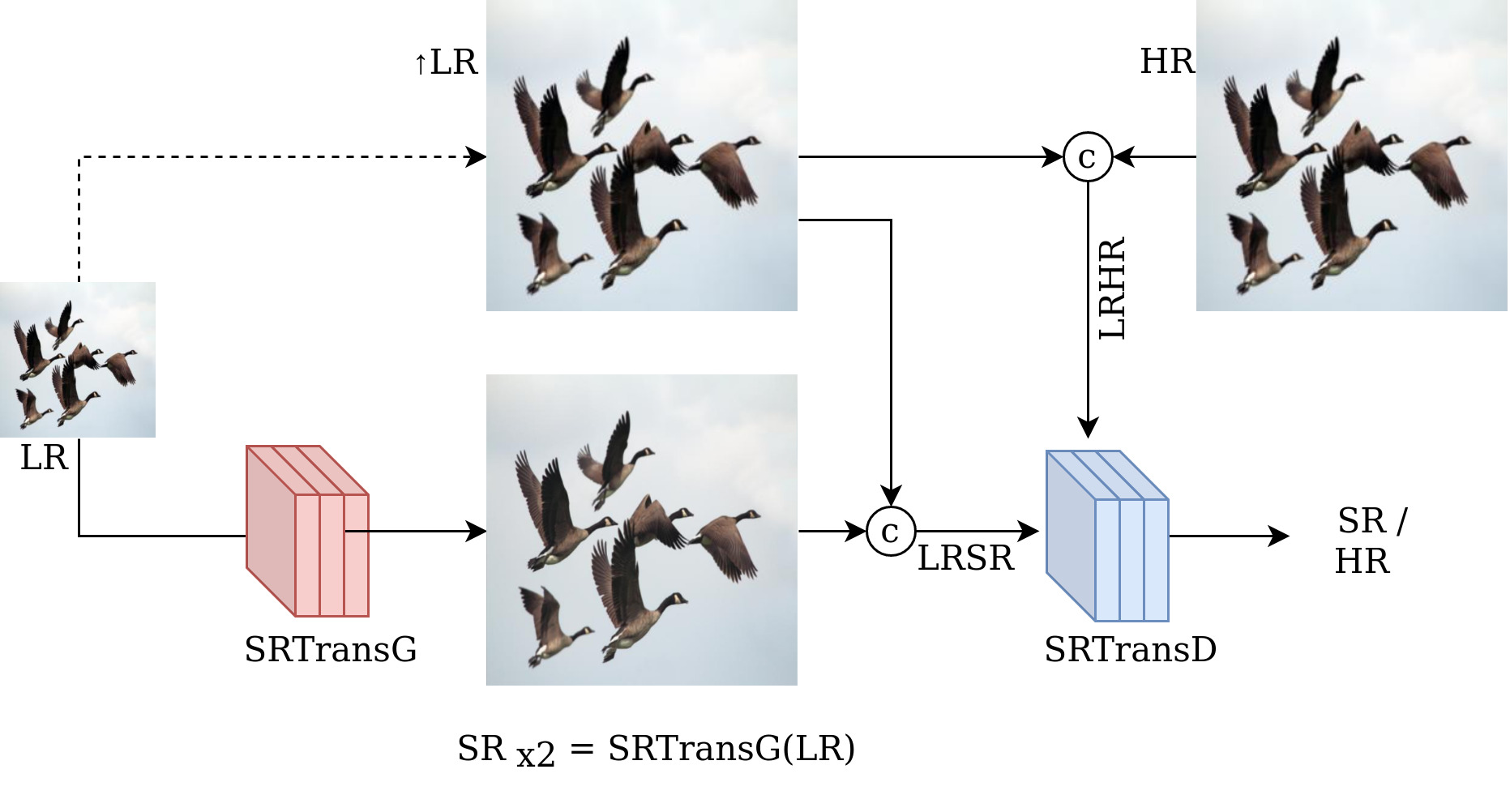}
\caption{Proposed SRTransGAN framework for image super-resolution.}
\label{fig:overview}
\end{figure}

Initially, the CNN based methods were proposed for image super-resolution. Super-Resolution Convolutional Neural Networks (SRCNN) \cite{SRCNNr1} was the initial attempt to use the CNN model for single image super-resolution (SISR) by utilizing three convolution layers. The error between the super-resolution images and ground truth high-resolution images was used to train the SRCNN model. This method has been tested to generate the super-resolution images at $2\times$, $3\times$ and $4\times$ scale. The SRCNN method required the Bicubic interpolation initially as a pre-processing step. However, Fast Super-Resolution Convolutional Neural Networks (FSRCNN) considered the low-resolution image directly as the input to CNN model and used Deconvolution layer \cite{FSRCNNr2}. The FSRCNN was faster than SRCNN, but the performance of FSRCNN was not satisfactory for super-resolution at higher scale. Very Deep Super-Resolution (VDSR) \cite{VDSRr3} considered a deep CNN for image super-resolution. In order to cope up the problem of training unstability, the VDSR used skip connection and gradient clipping strategy. 
Several CNN based image super-resolution methods also exploited the residual concept for improved training of the model, including Enhanced Deep Super-Resolution Network (EDSR) \cite{EDSR-baseline}, Residual Dense Network \cite{RDN}, Deep Recursive Residual Network (DRRN) \cite{DRRN}, Texture and Detail-Preserving Network (TDPN) \cite{tdpn}, Wavelet-Based Texture Reformation Network \cite{WBTRN}, Multi-Path Residual Network \cite{MPRN} and Cascading Residual Network (CARN) \cite{CARN}. 
Some researchers also tried to investigate the lightweight CNNs models for efficient image super-resolution, such as lightweight Information Multi-Distillation Network (IMDN) \cite{IMDN} and Cascading mechanism based image super-resolution \cite{CARN}. Other prominent CNN based image super-resolution approaches include Deep Laplacian Pyramid Networks (LapSRN) \cite{LapSRN}, Deeply-Recursive Convolutional Network (DRCN) \cite{DRCN}, Convolutional Super-Resolution Network for Multiple Noise-Free Degradations (SRMDNF) \cite{SRMDNF}, Residual Non-local Attention Network RNAN \cite{RNAN}, Ordinary Differential Equation (ODE)-inspired CNN for Image Super-resolution (OISR-RK3) \cite{OISR-RK3}, TV-TV Minimization \cite{robust_tvtv}, and Super-Resolution With Graph Attention Network (SRGAT)
\cite{srgat}. Zhou et al. \cite{IGNN} developed a cross-scale Internal Graph Neural Network (IGNN) for image super-resolution by exploiting the cross-scale patch recurrence property in images. 
Although all these models utilize the customized CNNs, the performance is comprosised due to lack of effective objective functions. Moreover, these models are unable to focus on high frequency regions and blur regions which are important regions requiring more attention. 

Generative adversarial networks (GANs) have been used successfully with outstanding performance for image-to-image translation task \cite{PIX, pcsgan} to transform the images from source to target domain. GANs contain generator and discriminator networks. 
The GAN based approaches have also been exploited for image super-resolution task \cite{SRGAN,esrgan,DUSGAN,GMGAN}. Among the initial approaches, Super-resolution GAN (SRGAN) \cite{SRGAN} utilized a CNN based generator and discriminator networks. Later, Enhanced Super-Resolution GAN (ESRGAN) \cite{esrgan} improved the network architecture, adversarial loss \& perceptual loss by using Residual Dense Block \& features before activation. Direct Unsupervised Super-Resolution GAN (DUS-GAN) \cite{DUSGAN} improved the perceptual quality with a novel mean opinion score as quality loss. GAN-Based Image Super-Resolution (GMGAN) \cite{GMGAN} improved training of GANs with a novel quality loss for image super-resolution task.
Though GAN based models have shown very promising image super-resolution but, their performance is limited by the learning capability of generator and discriminator networks.

The breakthrough in Transformer networks \cite{attention2017vaswani} has led to development of several transformer based models for different computer vision and image processing tasks \cite{khan2021transformers,vit, fan2021multiscale}. Recently, the transformers are also exploited for image super-resolution \cite{SAN, RCAN, ESRT, IPT}.
The transformer with CNN structure is used in ESRT \cite{ESRT}. It combines a transformer based backbone network after a convolutional network based backbone sequentially. This model uses lightweight backbone networks to make it computationally efficient.
Residual channel attention network (RCAN) \cite{RCAN} implements residual in residual block along with long skip connections to form a deep network for image super-resolution.
Pre-trained image processing transformer (IPT) \cite{IPT} uses a pre-trained model with ImageNet benchmark. IPT model contains multi-head and multi-tail networks with transformer encoder-decoder structure. This work reports high PSNR for different image processing applications, including image super-resolution. Though the researchers have explored the transformer models for image super-resolution, their performance is limited as they have not exploited the generative capabilities such as adversarial training. However, the recent trend suggests that the transformer network can greatly improve the performance of GAN models, such as ViTGAN \cite{vitgan} and TransGAN \cite{Transgan}.

\begin{figure*}[!t]
\centering
\includegraphics[width=\textwidth
]{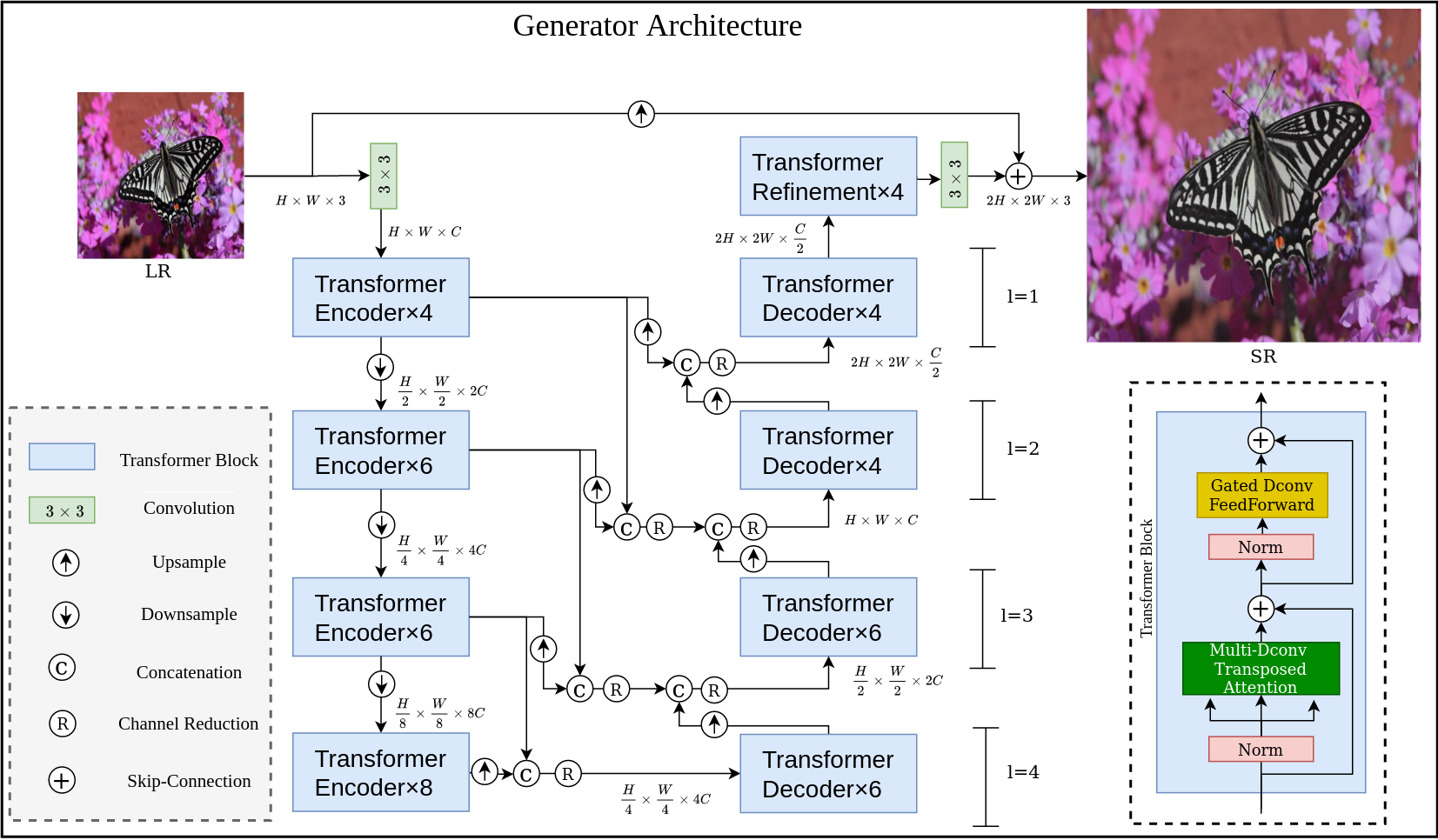}
\caption{Proposed Super-Resolution Transformer Generator (SRTransG) Network used in SRTransGAN.}
\label{fig:modelg}
\end{figure*}

Hence, motivated from the success of transformer models and GAN models, the objective of the proposed method is to develop a transformer based GAN (i.e., SRTransGAN) model for image super-resolution as shown in Fig. \ref{fig:overview}. In contrast to the existing transformer based image super-resolution methods, the proposed SRTransGAN is a conditional generative model and can learn the distribution of low-resolution and high-resolution images along with their mapping using adversarial training. 
The major contributions of the proposed work are as follows:
\begin{enumerate}
\item In this paper, we propose a new transformer based GAN model, namely SRTransGAN, for single image super-resolution by exploiting a transformer based generator network and a vision transformer based discriminator network.
\item The proposed transformer based generator network is based on the skip connection and follows a multi-level encoder-decoder structure with attention modules and concatenates the features at different scales by utilizing the down/up-sampling to learn the important features which is useful to reconstruct the pixels in high-resolution with better details of images.
\item We utilize the vision transformer based discriminator network as the vision transformers considers the images as sequence of patches and useful for classification problem. The adversarial training facilitated by the vision transformer based discriminator network helps the transformer based generator network to learn the distributions and mapping between low and high-resolution.
\item We conduct the extensive experiments on various image super-resolution datasets and compare the results with state-of-the-art models, including transformer ones, to show the effectiveness of the proposed model. We also perform the ablation study on different parameters like impact of different number of transformer blocks, levels and training datasets.
\end{enumerate} 

The proposed SRTransGAN model is detailed in Section \ref{sec:Proposed_SRTransGan_Method}, including overall structure, Generator and Discriminator architectures. The experimental settings, such as the datasets, implementation details and loss functions are summarized in Section \ref{sec:Experimental_Setup}. The experimental results and analysis are discussed in Section \ref{sec:Experimental_Results}. Ablation study is conducted in Section \ref{sec:Ablation_Study}. The conclusion is provided in Section \ref{sec:Conclusion}.

\section{Proposed SRTransGan Method} \label{sec:Proposed_SRTransGan_Method}
In this section, the overall structure of the proposed image super-resolution using transformer based GAN (SRTransGAN) is described. The proposed SRTransGAN framework is illustrated in Fig. \ref{fig:overview} which uses a generator network termed as SRTransG and a discriminator network termed as SRTransD. The SRTransG is based on the transformer \cite{restormer2021zamir} and generates the super-resolution (SR) image at $2\times$ dimension given a low-resolution (LR) image as input. The SRTransD is based on the vision transformer \cite{vitgan} which distinguishes the real high-resolution (HR) images from the generated super-resolution (SR) images. Both the HR and SR images are first channel wise concatenated with up-sampled version of LR image and then used as input to the SRTransD network.

\subsection{Transformer Generator}
A low-resolution (LR) image is transformed into super-resolution (SR) image using generator network. The proposed SRTransGAN uses Transformer-based SRTransG generator network. 
The proposed SRTransG network generates the SR images at increasing scales in a progressive manner. Basivally, a down-sampling is performed using the Transformer Encoder modules and then up-sampling is performed using the Transformer Decoder modules as illustrated in Fig. \ref{fig:modelg}. The Transformer Encoder and Transformer Decoder modules exploit the global relationship in a better way apart from the local relationship which is very essential to synthesize the images at higher-resolution. The SRTransG generates the super-resolution images at $2\times$ scale. In order to generate the SR images at $4\times$ scale, we repeat the SRTransG twice with same parameters. 

\paragraph{Overall Pipeline}
The SRTransG network receives input as a low-resolution image $LR \in \mathbb{R}^{H,W,3}$, where $H$ is the height, $W$ is the width and $3$ is the color channels of the input image. First we apply a convolution operation to obtain the low-level feature embeddings of overlapping patches, represented as $F_0 \in \mathbb{R}^{H,W,C}$, where $C$ is the number of channels in the feature map. These low-level features $F_0$ pass through a transformer encoder-decoder structure and transform the $F_0$  into $F_8 \in \mathbb{R}^{2H,2H,C/2}$. The encoder-decoder structure has $l$ levels of hierarchy. We use $l=4$ in the experiments and also perform the analysis for different levels. At each level $l$ there is a transformer encoder $E^l_n$ and transformer decoder $D^l_n$ block, where $n$ is the number of transformer stack. The transformer encoder at $l=1$ is $E^1_4$ processes the low level features $F_0 \in \mathbb{R}^{H,W,C}$ and transforms to $F_1 \in \mathbb{R}^{H,W,C}$. It is represented as,
\begin{equation}
F_1 = E^1_4(F_0).
\end{equation}

At level $l=2$ the transformer encoder $E^2_6$ processes the down-sampled feature $\downarrow$$F_1 \in \mathbb{R}^{H/2,W/2,2C}$ to $F_2 \in \mathbb{R}^{H/2,W/2,2C}$ as follows,
\begin{equation}
F_2 = E^2_6(\downarrow F_1)
\end{equation}
where $\downarrow$ denotes the down-sampling to half resolution and double number of channels using the convolution operation.

At level $l=3$ the transformer encoder $E^3_6$ processes the down-sampled feature $\downarrow$$F_2 \in \mathbb{R}^{H/4,W/4,4C}$ to $F_3 \in \mathbb{R}^{H/4,W/4,4C}$ as follows,
\begin{equation}
F_3 = E^3_6(\downarrow F_2).
\end{equation}

At level $l=4$ the transformer encoder $E^4_8$ processes the down-sampled feature $\downarrow$$F_3 \in \mathbb{R}^{H/8,W/8,8C}$ to $F_4 \in \mathbb{R}^{H/8,W/8,8C}$ as follows,
\begin{equation}
F_4 = E^4_8(\downarrow F_3).
\end{equation}

At level $l=4$ the transformer decoder $D^4_6$ processes the concatenated feature ($\in \mathbb{R}^{H/4,W/4,4C}$) of up-sampled feature $\uparrow$$F_4$ and $F_3$ with channel reduction to $F_5 \in \mathbb{R}^{H/4,W/4,4C}$ as follows, 
\begin{equation}
F_5 = D^4_6(\circled{R}((\uparrow F_4)  \circled{c} (F_3)))
\end{equation}
where $\circled{R}(.)$ is used to reduce the number of channels by half using convolution operation,  $(.)\circled{c}(.)$ is the channel-wise concatenation operation  and $\uparrow$ denotes the up-sampling to double spatial resolution and half number of channels using convolution operation. As the concatenation operation increases the number of channel, therefore to maintain the consistency at different levels, we use the reduction operation after every concatenation operation.

At level $l=3$ the concatenated feature of up-sampled features $\uparrow$$F_3$ and $F_2$ with channel reduction is further concatenated with the up-sampled features $\uparrow$$F_5$ with channel reduction and processed by transformer decoder $D^3_6$ to $F_6 \in \mathbb{R}^{H/2,W/2,2C}$ as follows, 
\begin{equation}
F_6 = D^3_6(\circled{R}((\uparrow F_5)  \circled{c} (\circled{R}((\uparrow F_3)  \circled{c}(F_2) ))).
\end{equation}

At level $l=2$ the concatenated feature of up-sampled features $\uparrow$$F_2$ and $F_1$ with channel reduction and this feature is further concatenated with of up-sampled features $\uparrow$$F_6$ with channel reduction is processed by transformer decoder $D^2_4$ to $F_7 \in \mathbb{R}^{H,W,C}$ as follows, 

\begin{equation}
F_7 = D^2_4(\circled{R}((\uparrow F_6)  \circled{c} (\circled{R}((\uparrow F_2)  \circled{c} (F_1))))
\end{equation}

At level $l=1$ the transformer decoder  $D^1_4$ processes the concatenated feature of up-sampled feature $\uparrow$$F_1$ and $\uparrow$$F_7$ with channel reduction to $F_8 \in \mathbb{R}^{2H,2W,C/2}$ as follows,

\begin{equation}
F_8 = D^1_4(\circled{R}((\uparrow F_1)  \circled{c} (\uparrow F_7))).
\end{equation}

After this encoder-decoder structure one another transformer refinement block is applied to refine the $F_8$ high-level features to $F_9$. This $F_9$ is passed with a convolution layer to obtain an image in high resolution $\in \mathbb{R}^{2H,2W,C}$. The Skip connection is applied between the generated feature image and up-sampled LR to generate a super-resolution image SR $\in \mathbb{R}^{2H,2W,C}$. It helps the model to learn the difference mapping between the LR and SR images.

\paragraph{Transformer block}
In the transformer encoder, transformer decoder and transformer refinement block architectures, we exploit a Multi Deconvolution (Dconv) transposed attention by following \cite{restormer2021zamir} which applies self attention over the channel. 
Before Multi Dconv transposed attention a normalised layer $Y \in \mathbb{R}^{H,W,C}$ is used. In  Multi Dconv transposed attention \cite{restormer2021zamir}, query (Q), key (K) and value (V) projections are computed by combining the pixel-wise cross-channel context with $1 \times 1$ convolution $W^{(.)}_p$ and depth-wise channel context with $3\times3$ convolution $W^{(.)}_p$, i.e., $Q=W^Q_d W^Q_ p Y$, $K=W^K_d W^K_p Y$ and $V=W^V_d W^V_p Y$.
Then query and key projections are reshaped to generate a transposed-attention map $A \in \mathbb{R}^{C,C}$. 

\begin{figure}[!t]
\centering
\includegraphics[width=\columnwidth]{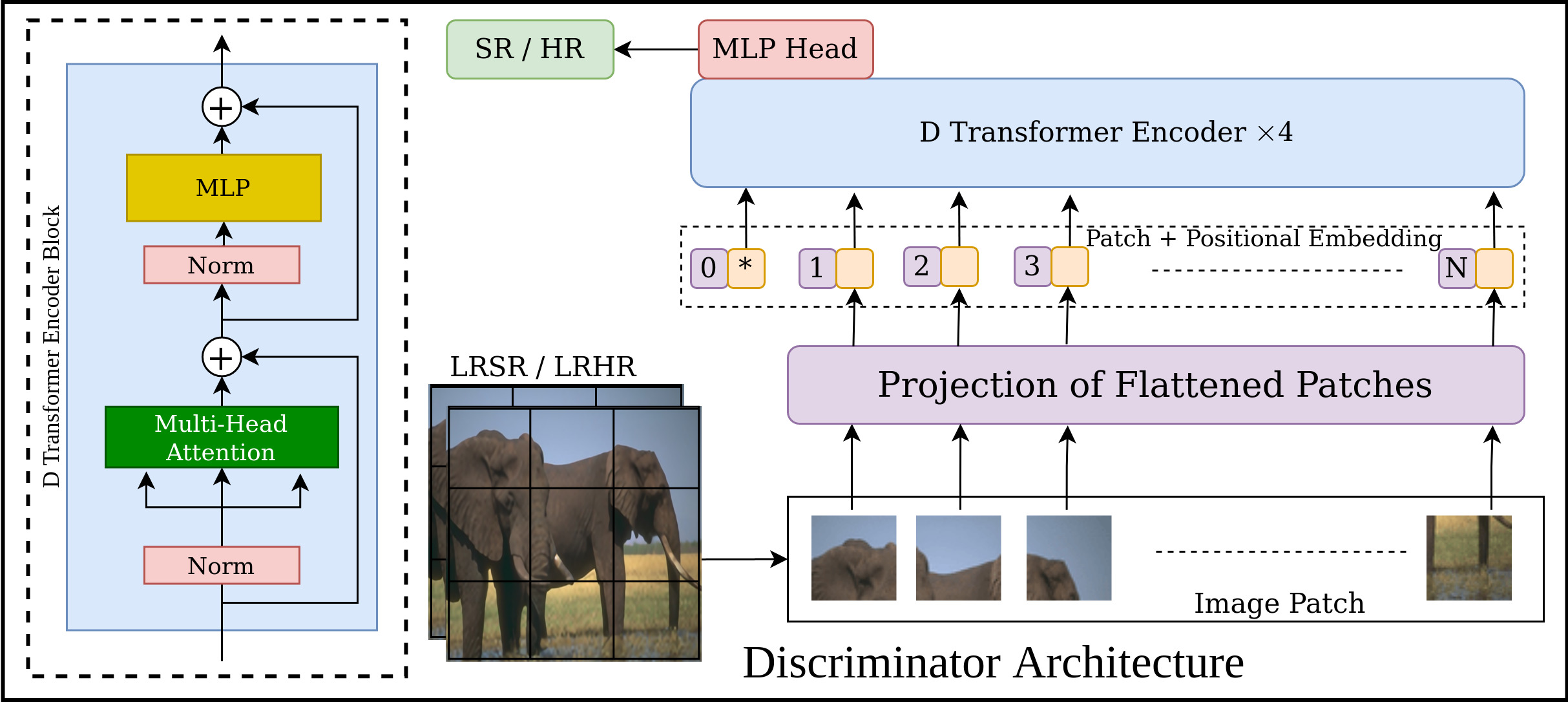}
\caption{The Discriminator Network (SRTransD) used in SRTransGAN model. It is based on the Vision Transformer.}
\label{fig:modeld}
\end{figure}

After Multi Dconv transposed attention, a skip connection is applied from the input of the transformer block. In addition to this, further a normalised layer is used before Gated Dconv-Feed Forward block similar to \cite{restormer2021zamir}. To transform features, two parallel $1 \times 1$ depth-wise convolutions are used followed by a $3 \times 3$ Dconv layer in the Gated Dconv-Feed Forward block. Finally, this block uses a gating mechanism to improve the feature transformation by computing the element-wise product of two parallel paths in which one is activated with the GELU non-linearity \cite{gelu}.  
Given an input tensor $X \in \mathbb{R}^{H,W,C}$, $\text{Gating}(X)$ is described as follows, 
\begin{equation}
\text{Gating}(X) = \Phi (W^1_d W^1_p (LN(X)))\odot W^2_d W^2_p (LN(X)) 
\end{equation}
where, LN is the layer normalization \cite{layerNORM}, $\Phi$ represents the GELU activation function, and $\odot$ denotes element-wise multiplication.

\subsection{Transformer Discriminator}
Similar to ViTGAN \cite{vitgan}, the proposed SRTransGAN model uses the Transformer Discriminator (SRTransD), which is based on the ViT \cite{vit}. Using ViT-based discriminator matches the complexity of Transformer-based generator in the proposed model. The SRTransD network is illustrated in Fig. \ref{fig:modeld}. Next, we explain the different modules of SRTransD.  

\paragraph{Patch Module}
The $N$ overlapping image patches $I_i  \in \mathbb{R}^{k,k,c}$ are generated from the input image $I \in \mathbb{R} ^{m,m,c}$, where $i = 1,2,3,... N$, $k$ is the spatial resolution of patches and $c$ is the number of channels. We use the overlapping patches to avoid the information loss at borders of the patches and better utilize the continuity in the information between the patches. Basically, a dynamic stride length $S_h,S_w$ is taken so that patches have a continuous flow of adjacent information. A patch of dimension $I_i \in \mathbb{R}^{k,k,c}$ is converted into a linear vector $V_i  \in \mathbb{R}^{1,d}$  by flattening as the transformer requires vectors as input. 

\paragraph{Class Token and Positional Embedding Module}
First an embedding of dimension $de$ is generated for each patch from its flattened vector ($RV_i \in \mathbb{R}^{1,d}$ for $i^{th}$ patch) as follows,
\begin{equation}
PE_i = RV_i \times W_{PE}
\end{equation}
where $W_{PE}$ is parameter and  $W_{PE} \in \mathbb{R}^{d,de}$ is the parameter matrix and $PE_i \in \mathbb{R}^{1,de}$ is the embedding w.r.t. $i^{th}$ patch.
An additional class token embedding $ CT \in  \mathbb{R}^{1,de}$ is generated randomly.  To generate the expanded embedding $(EE \in \mathbb{R}^{N+1,de})$ the class token CT is concatenated with the projected embedding PE as EE = [CT, PE] on $1^{st}$ dimension. The positional embedding $(PoE \in \mathbb{R}^{N+1,de})$ is included with zero initialized as the learn-able parameters and added in the expanded embedding in order to incorporate the spatial information. Therefore the final projection embedding (FE) as follows,
\begin{equation}
FE = dropout(EE + PoE)
\end{equation}
where $dropout$ is a dropout layer \cite{Dropout} and $FE \in \mathbb{R}^{N +1,de}$ is the final projection embedding which is passed as input to transformer encoder module.

\paragraph{Transformer Encoder Module}
The transformer encoder module consists the stack of L transformer blocks to increase the hierarchical learning capacity. A transformer block transforms the given input embedding $h_j \in \mathbb{R}^{ N+1,de}$ to an output embedding having same dimension using multi-head self-attention mechanism. For the transformer block $T_j|_{j=1,2,3,...L}$ input will be hidden state $h_{j-1}$ obtained form previous transformer block $T_{j-1}$. Input for the first transformer block $T_1$ will be the final projection embedding, i.e., $h_0 = FE$. Output of the final transformer block $T_L$ will be $h_L$. In the transformer block $T_j$ linear normalization is used as first layer which takes $h_{j-1}$ as input and produces $l_{j-1} \in \mathbb{R}^{N +1,de}$ as output. The output $l_{j-1}$ is given as input to the multi-head self-attention which is made up of three parametric projections as Query, Key and Value. The output of linear projections are
\begin{equation}
S_t = l_{j-1} \times W_t
\end{equation}
here $S_t \in \mathbb{R}^{N+1,de}$ and $W_t$ are the trainable weights for $t = \{Q, K, V\}$. 
The attention weights ($A_w$) are calculated as:
\begin{equation}
A_w = softmax(\frac{S_Q \times S_K}{\sqrt{A_S}}), 
\end{equation}
where $A_S = de/A_h$.
Attention weighted features ($F_A \in \mathbb{R}^{A_h,N +1,A_S}$) are computed from the attention weights ($A_w$) and Value vector ($S_V$) as follows,
\begin{equation}
F_A = A_w \times S_V.
\end{equation}
Attention weighted features $F_A \in \mathbb{R}^{A_h,N +1,A_S}$ is reshaped to $F_A \in \mathbb{R}^{N +1,de}$. Finally, a linear projection is applied to $F_A$ with a learnable parameter $W_A$ producing $F_S = F_A \times W_A$ as the output of the self-attention module ($F_S$). The residual connection is included in the transformer blocks to optimise the training. 

\begin{equation}
F_R = h_{j-1} + F_S.
\end{equation}
The output of residual connection $F_R$ is then normalised with the output of this layer as $l_R$  of the same dimensions. A multi-layer perceptron module is applied on the output of normalization with a linear projection to $\text{mlp}_{dim} = de \times \text{mlp}_{ratio}$ dimension, GELU activation function and a linear projection back to $de$ dimension. Finally, a residual connection is employed to generate the transformer block's output as, 
\begin{equation}
h_j = F_R + F_{MLP}
\end{equation}
where, $F_{MLP} \in \mathbb{R}^{N+1,de}$ is the MLP module's output, $F_R$ the first residual connection's output, and $h_j$ is the output of the $j^{th}$ transformer block. The layer normalisation output on last transformer block $h_L$ is transferred to the following step by the transformer encoder's final output ($T_E \in \mathbb{R}^{N +1,de}$). A multi-layer perceptron (MLP) head to compute the probability of input image being classified into real high-resolution (HR) category.

\begin{table}[t]
\caption{PSNR and SSIM comparision has been performed among diffrent SR methords on diffrent dataset and diffrent scale ($2\times$  and $4\times$  super-resolution)}
\label{tab:PSNR/SSIMx2x4}
\resizebox{\columnwidth}{!}{
\begin{tabular}{m{0.20\columnwidth}m{0.022\columnwidth}m{0.13\columnwidth}m{0.13\columnwidth}m{0.13\columnwidth}m{0.13\columnwidth}}
\hline
Method &Sca & Set5 & Set14 & BSD100 & Urban100 \\
&le& PSNR/SSIM & PSNR/SSIM & PSNR/SSIM & PSNR/SSIM \\
\hline
SRCNN \cite{SRCNNr1} & &  36.66/0.9542 & 32.45/0.9067 & 31.36/0.8879 & 29.50/0.8946 \\
FSRCNN \cite{FSRCNNr2} & &37.00/0.9558 & 32.63/0.9088 & 31.53/0.8920 & 29.88/0.9020 \\
VDSR \cite{VDSRr3} & &  37.53/0.9587 & 33.03/0.9124 & 31.90/0.8960 & 30.76/0.9140  \\
DRCN \cite{DRCN} & & 37.63/0.9588 & 33.04/0.9118 & 31.85/0.8942 & 30.75/0.9133  \\
LapSRN \cite{LapSRN}& & 37.52/0.9591 & 32.99/0.9124 & 31.80/0.8952 & 30.41/0.9103 \\
DRRN \cite{DRRN} & & 37.74/0.9591 & 33.23/0.9136 & 32.05/0.8973 & 31.23/0.9188 \\
MemNet \cite{Memnet} & & 37.78/0.9597 & 33.28/0.9142 & 32.08/0.8978 & 31.31/0.9195 \\
EDSR \cite{EDSR-baseline} & & 37.99/0.9604 & 33.57/0.9175 & 32.16/0.8994 & 31.98/0.9272  \\
SRMDNF \cite{SRMDNF} & & 37.79/0.9601 & 33.32/0.9159 & 32.05/0.8985 & 31.33/0.9204  \\
CARN \cite{CARN} & $2\times$ & 37.76/0.9590 & 33.52/0.9166 & 32.09/0.8978 & 31.92/0.9256  \\
IMDN \cite{IMDN} & & 38.00/0.9605 & 33.63/0.9177 & 32.19/0.8996 & 32.17/0.9283  \\
ESRT \cite{ESRT} & & 38.03/0.9600 & 33.75/0.9184 & 32.25/0.9001 & 32.58/0.9318  \\
RCAN \cite{RCAN} & & 38.27/0.9614 & 34.12/0.9216 & 32.41/0.9027 & \underline{33.34/0.9384}  \\
OISR \cite{OISR-RK3} & & 38.21/0.9612 & 33.94/0.9206 & 32.36/0.9019 & 33.03/0.9365  \\
RNAN \cite{RNAN} & & 38.17/0.9611 & 33.87/0.9207 & 32.32/0.9014 & 32.73/0.9340  \\
SAN \cite{SAN} & & 38.31/\underline{0.9620} & 34.07/0.9213 & 32.42/\underline{0.9028} & 33.10/0.9370 \\
IGNN \cite{IGNN} & & 38.24/0.9613 & 34.07/\underline{0.9217} & 32.41/0.9025 & 33.23/0.9383 \\
IPT \cite{IPT} &  & \underline{38.37}/0.959$^*$ & \underline{34.43}/0.924$^*$ & \underline{32.48}/0.943$^*$ & \textbf{33.76/0.9535}$^*$  \\
Proposed & & \textbf{43.862}/\textbf{0.986} & \textbf{36.162}/\textbf{0.944} & \textbf{36.977}/\textbf{0.957} & 31.935/0.933 \\
\hline
SRCNN \cite{SRCNNr1} & & 30.48/0.8628 & 27.50/0.7513 & 26.9/0.7101 & 24.52/0.7221 \\
FSRCNN \cite{FSRCNNr2} & & 30.72/0.8660 & 27.61/0.7550 & 26.98/0.7150 & 24.62/0.7280 \\
VDSR \cite{VDSRr3} & &  31.35/0.8838 & 28.01/0.7674 & 27.29/0.7251 & 25.18/0.7524  \\
DRCN \cite{DRCN} & & 31.53/0.8854 & 28.02/0.7670 & 27.23/0.7233 & 25.14/0.7510\\
LapSRN \cite{LapSRN} & & 31.54/0.8852 & 28.09/0.7700 & 27.32/0.7275 & 25.21/0.7562 \\
DRRN \cite{DRRN} & & 31.68/0.8888 & 28.21/0.7720 & 27.38/0.7284 & 25.44/0.7638  \\
MemNet \cite{Memnet} & & 31.74/0.8893 & 28.26/0.7723 & 27.40/0.7281 & 25.50/0.7630  \\
EDSR \cite{EDSR-baseline} & & 32.09/0.8938 & 28.58/0.7813 & 27.57/0.7357 & 26.04/0.7849  \\
SRMDNF \cite{SRMDNF} & & 31.96/0.8925 & 28.35/0.7787 & 27.49/0.7337 & 25.68/0.7731  \\
CARN \cite{CARN} & $4\times$ & 32.13/0.8937 & 28.60/0.7806 & 27.58/0.7349 & 26.07/0.7837  \\
IMDN \cite{IMDN} & &  32.21/0.8948 & 28.58/0.7811 & 27.56/0.7353 & 26.04/0.7838  \\
ESRT \cite{ESRT} & & 32.19/0.8947 & 28.69/0.7833 & 27.69/0.7379 & 26.39/0.7962  \\
RCAN \cite{RCAN} & & 32.63/0.9002 & 28.87/0.7889 & 27.77/\underline{0.7436} & 26.82/\underline{0.8087}  \\
OISR \cite{OISR-RK3} & & 32.53/0.8992 & 28.86/0.7878 & 27.75/0.7428 & 26.79/0.8068  \\
RNAN \cite{RNAN} & &  32.49/0.8982 & 28.83/0.7878 & 27.72/0.7421 & 26.61/0.8023 \\
SAN \cite{SAN} & &  \underline{32.64}/\underline{0.9003} & 28.92/0.7888 & 27.78/\underline{0.7436} & 26.79/0.8068 \\
IGNN \cite{IGNN} & & 32.57/0.8998 & 28.85/\underline{0.7891} & 27.77/0.7434 & \underline{26.84}/\textbf{0.8090} \\
IPT \cite{IPT} & & \underline{32.64}/0.8260$^*$ & \underline{29.01}/0.6783$^*$ & \underline{27.82}/0.6800$^*$ & \textbf{27.26}/0.7952$^*$ \\
Proposed & & \textbf{36.941}/\textbf{0.944} & \textbf{29.509}/\textbf{0.828} & \textbf{30.322}/\textbf{0.823} & 25.519/0.761  \\ \hline
\multicolumn{6}{c} {Here, $*$ denotes the reproduced results from pre-train model.}
\end{tabular}}
\end{table}


\begin{table}[]
\centering
\caption{PSNR and SSIM comparison has been performed among different FSR methods ($4\times$ scale face super-resolution) on CelebA Dataset}
\label{tab:PSNR/SSIM_FSR}
\begin{tabular}{|m{0.23\columnwidth}|m{0.1\columnwidth}|m{0.1\columnwidth}|m{0.1\columnwidth}|}
\hline
\multirow{2}{*}{Method} & \multicolumn{3}{c|}{ 4× ($128^2->512^2$)}  \\ \cline{2-4}
 & $PSNR\uparrow$ & $SSIM\uparrow$ & $FID\downarrow$ \\ 
 \hline
GFPGAN \cite{GFPGAN} & 27.32 & 0.7686 & 36.76\\
GPEN \cite{GPEN}  & 27.10 & 0.7593 & 43.81\\
HiFaceGAN \cite{HiFaceGAN} & 27.39 & 0.7397  & 30.28\\
DFDNet \cite{DFDNet} & 26.47 & \underline{0.7802}  & 45.80 \\
PSFRGAN \cite{PSFRGAN} & 27.24 & 0.7607 & 35.85\\
GCFSR \cite{GCFSR} & \underline{27.81} & 0.7711 & \textbf{27.90}\\ \hline
Proposed & \textbf{35.12} & \textbf{0.7876} & 33.67 \\ \hline
\end{tabular}
\end{table}

\section{Experimental Setup} \label{sec:Experimental_Setup}
In this section, the experimental settings are presented such as datasets, performance measures, implementation details and loss functions.

\subsection{Datasets and Metrics}
The Proposed model is trained and tested with the DIV2K dataset \cite{datadiv} for SISR problem. It contains 900 high-resolution images with its corresponding low-resolution images. Out of which, 800 and 100 image pairs are used in the DIV2K training set and validation set, respectively. These images are rich with textures and objects. The high-resolution images are available at different scales such as $2\times$, $3\times$ and $4\times$. The SRTransGAN is also evaluated on Set5 \cite{set5}, Set14 \cite{set14}, BSD100 \cite{bsd100} and Urban100 \cite{urban100} datasets. 
The Set5 dataset contains 5 images for testing the model's performance on super-resolution problem. The Set14 dataset contains 14 commonly used images for testing super-resolution performance of different models. BSD100 dataset contains 100 classical test images.
The Urban100 dataset consists of 100 images of urban scenes in super-resolution test set.
The CUFED \cite{datacufed} dataset is also used in the ablation study to show the impact of training dataset. 
The proposed model is also trained on FFHQ Dataset consisting of 70K images for FSR problem. The proposed model achieves a very good performance for Face Super-resolution also. For FSR problem, the proposed work is also evaluated on FFHQ test dataset and CelebA test set. The 202,599 face images of 10,177 celebrities are present in CelebA dataset.
The PSNR and SSIM metrics are used for evaluation. These metrics are computed between the ground truth high-resolution images and the generated super-resolution images. 

\subsection{Implementation Details}
We train our SRTransGAN model on $LR \in \mathbb{R}^{64,64,3}$ low-resolution images to $HR \in \mathbb{R}^{128,128,3}$ super-resolution transformation (i.e., at $2 \times$ scale). 
Both Generator $SRTransG$ and discriminator $SRTransD$ networks have different transformer blocks. We use 4 levels in the proposed $SRTransG$ with [4,6,6,8] number of transformer stacks, respectively. Note that the number of parameters at a given level depends upon the number of transformer stack as that level with more parameters for more number of transformer stack. It helps the model to have the more learning ability at higher levels, i.e., more abstract features. While, in $SRTransD$, the number of transformer stacks are set to 4. In $SRTransG$, the kernel size is set to $3 \times 3$ in all convolution layers.
The learning rate is $2 \times 10^{-4}$. For up-scaling and down-scaling in $SRTransG$ convolution layer based module is applied. The up-scaling leads to double spatial resolution with half number of channels and the down-sampling leads to half spatial resolution with double number of channels. The generator module is used twice to generate the $4\times$ super-resolution images. The implementations are performed using Pytorch framework using different GPUs.

\subsection{Loss Function}\label{sec:Loss_Function}
In order to train the proposed SRTrasGAN model, a combination of two loss functions (i.e., Adversarial loss and Reconstruction loss) are used. Further, the Adversarial loss is a combination of Generator loss and Discriminator loss. The overall Generator loss can be give as:
\begin{equation}
 \mathcal{L}_{G} = arg \min_{G} \max_{D}\mathcal{L}_{adv} (G,D) \lambda_{adv}+  \mathcal{L}_{rec} \lambda_{rec}
\end{equation}
where, $\mathcal{L}_{adv}$ is the Adversarial loss, $\mathcal{L}_{rec}$ is the Reconstruction loss, and $\lambda_{adv}$ \& $\lambda_{rec}$ are the weight coefficients for Adversarial and Reconstruction loss, respectively.

Adversarial loss: Generative adversarial networks \cite{goodfellow2014generative} are able to generate high-quality images. We also use the Adversarial loss. The Generator and Discriminator losses are defined as,
\begin{equation}
\begin{multlined}
\mathcal{L}_{adv}(G,D) = \underset{SR \sim \mathbb{P}_g}{\mathbb{E}} [log(SRTransD(\uparrow LR \circled{c} HR))] + \\ \underset{SR \sim \mathbb{P}_g}{\mathbb{E}} [ log (1 - SRTransD(\uparrow LR \circled{c} SR)] 
\end{multlined}
\end{equation}
where $\uparrow$$LR$, $SR$, and $HR$ are the up-sampled low-resolution, super-resolution and high-resolution images, respectively, $\mathbb{P}_g$ and $\mathbb{P}_r$ are the distribution of generated super-resolution samples and high-resolution samples, respectively, and $\circled{c}$ is channel-wise concatenation. Note that the SR images are generated using generator as $SR = SRTransG(LR)$ and the $\uparrow$$LR$ is concatenated to $SR$ and $HR$ as the condition used in conditional GAN \cite{PIX}.

Reconstruction loss: The reconstruction loss is used to train the proposed generator network. It is a measure of pixel-level similarity between the the super-resolution and high-resolution images. This loss function is defined as:
\begin{equation}
\mathcal{L}_{rec} = \frac{1}{2H\times2W\times3} \left\| HR-SR \right\|_1.
\end{equation}
For SR methods adversarial loss provides better visual quality and reconstruction loss provides better pixel level generation. Therefore we train our model on both loss functions.

\begin{figure}[t]
\centering
\includegraphics[width=\columnwidth]{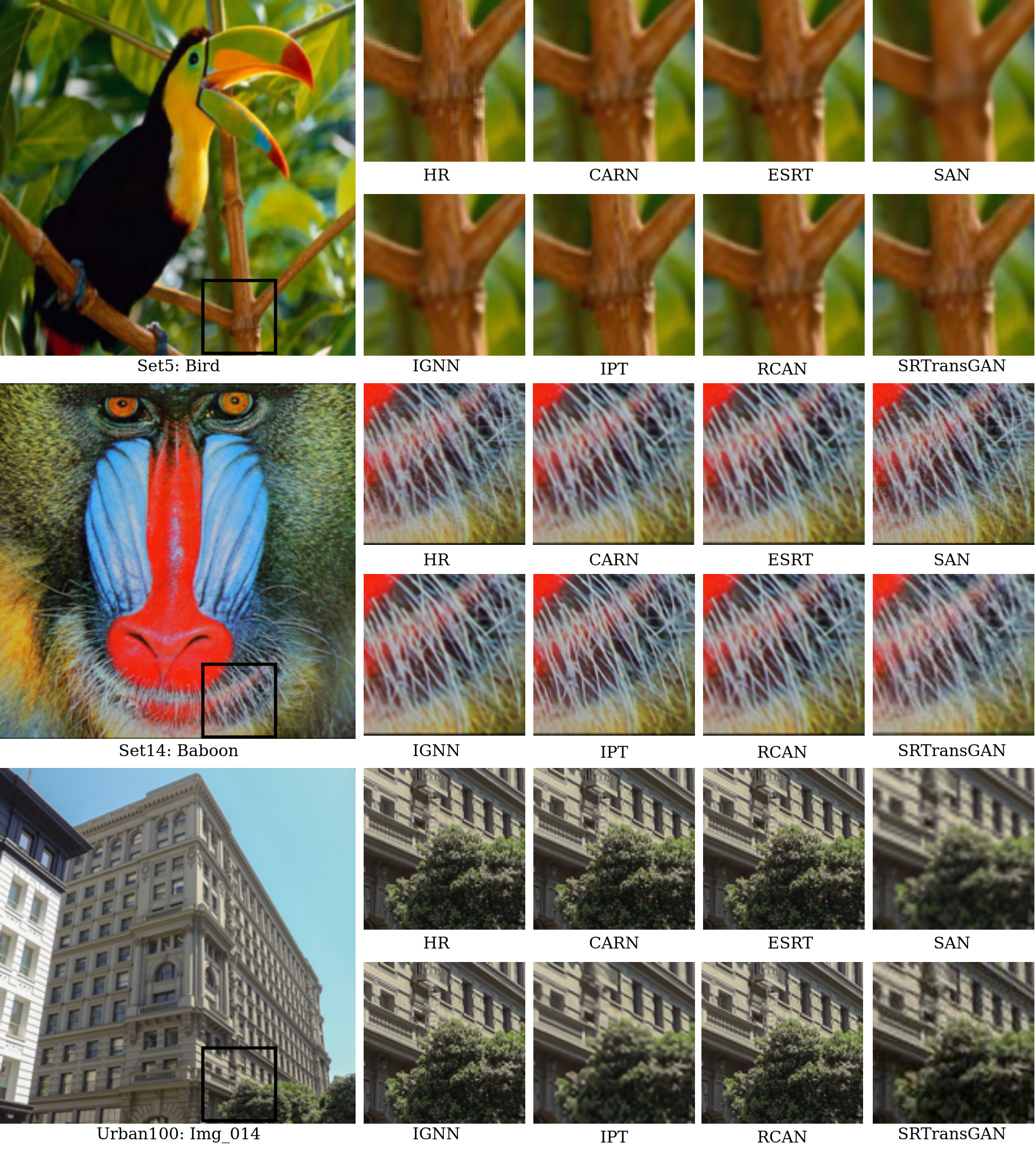}
\caption{Visualization results for $2\times$ super-resolution, (a) LR image, (b) SR image, and (C) HR image.}
\label{fig:datasetx2}
\end{figure}

\begin{figure}[t]
\centering
\includegraphics[width=\columnwidth]{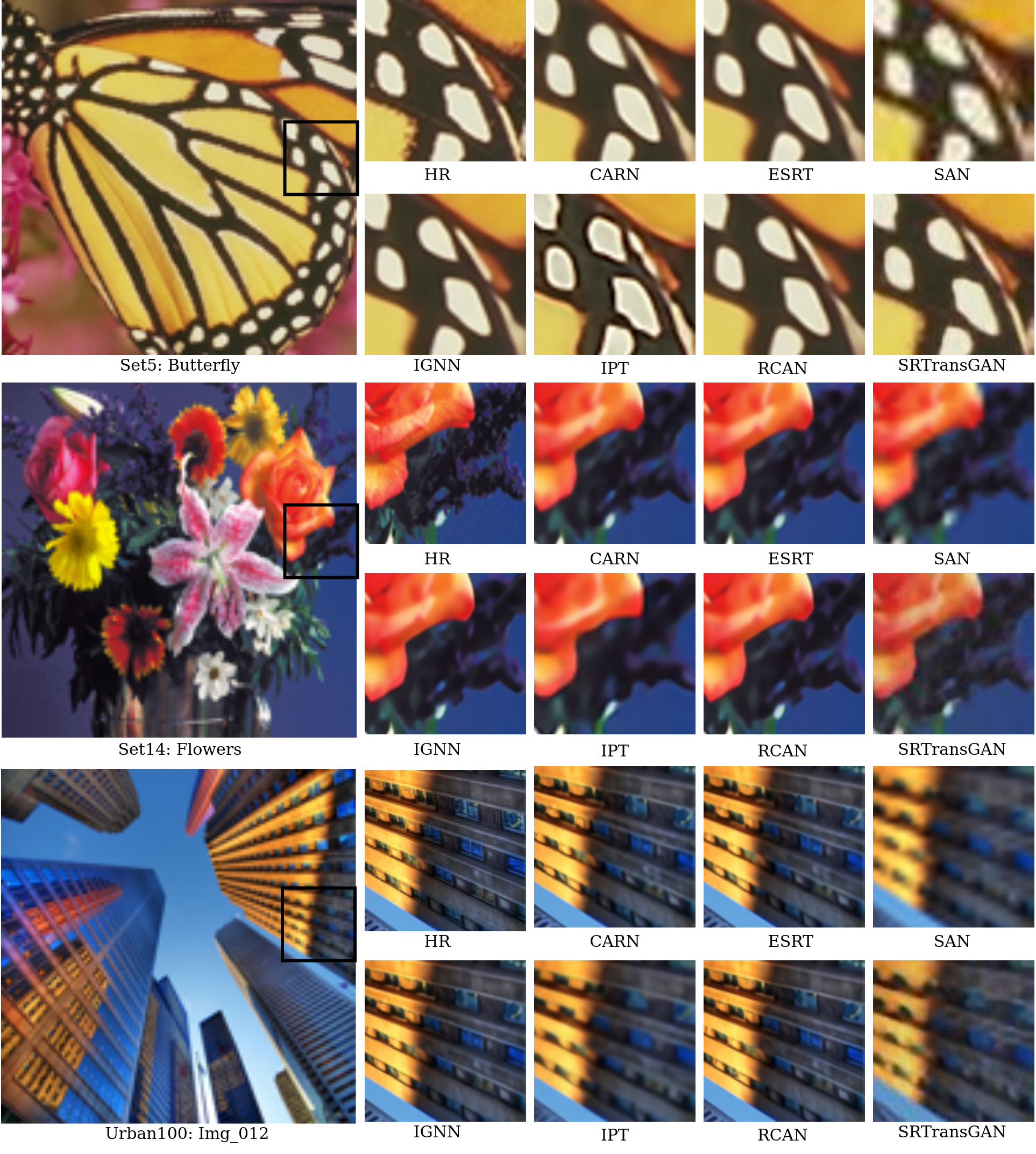}
\caption{Visualization results for $4\times$ super-resolution, (a) LR image, (b) SR image, and (C) HR image.}
\label{fig:datasetx4}
\end{figure}

\section{Experimental Results and Analysis} \label{sec:Experimental_Results}
In this section, we evaluate the performance of the proposed SRTransGAN model in various aspects like quantitative, qualitative and visual activation map along with the analysis. 
We also compare SRTransGAN model with the other state-of-art methods such as RCAN \cite{RCAN}, SAN \cite{SAN}, IGNN \cite{IGNN}, and IPT \cite{IPT}, among others.
All experiments are performed on $2\times$ and $4\times$ scale super-resolution between LR and HR images. We train our model on DIV2K training set and test on Set5, Set14, BSD100, and Urban100 datasets. Rather then this, we have trained and test our model for experimental settings for different super-resolution tasks: Face super-resolution. In this task we have trained our model on FFHQ training set and test on FFHQ and CelebA test set. We have also performed experiments varying number of transformer Stack, Levels in Encoder-Decoder Structure and Training Dataset is shown in Supplementary materials.

\subsection{Quantitative Evaluation: Single Image Super Resolution}
The proposed SRTransGAN method is compared with the state-of-the-art methods, such as SRCNN \cite{SRCNNr1}, FSRCNN \cite{FSRCNNr2}, VDSR \cite{VDSRr3}, DRCN \cite{DRCN}, LapSRN \cite{LapSRN}, DRRN \cite{DRRN}, MemNet \cite{Memnet}, EDSR \cite{EDSR-baseline}, SRMDNF \cite{SRMDNF}, CARN \cite{CARN}, IMDN \cite{IMDN}, ESRT \cite{ESRT}, RCAN \cite{RCAN}, OISR \cite{OISR-RK3}, RNAN \cite{RNAN}, SAN \cite{SAN}, IGNN \cite{IGNN}, IPT \cite{IPT}. Among these state-of-the-art models, the RCAN \cite{RCAN} has achieved higher performance in terms of both PSNR and SSIM in recent years for $2\times$ super-resolution on Urban100 dataset; IGNN \cite{IGNN} has achieved higher performance on $4\times$ super-resolution on Urban100; and IPT \cite{IPT} has achieved higher performance on Set5, Set14, BSD100, Urban100 in terms of PSNR metric.

The quantitative performance of the proposed SRTransGAN model for $2\times$ and  $4\times$ image super-resolution is shown in Table \ref{tab:PSNR/SSIMx2x4}. Here, bold text represents the highest results while underlined text represents the second highest results. 
For both $2\times$ and $4\times$ image super-resolution, all the models are outperformed by the proposed SRTransGAN model on Set5, Set14 and BSD100 datasets. Moreover, the comparable results are achieved on Urban100 dataset by SRTransGAN.
Basically, the test set of Urban100 dataset contains of particular urban scenes  while training set of DIV2K dataset contains images like buildings, animals, scene, etc.
The quantitative comparison results demonstrate the superiority of our proposed SRTransGAN over state-of-the-art SR approaches. 
    

\subsection{Quantitative Evaluation: Face Super Resolution}
The proposed SRTransGAN method is compared with the state-of-the-art methods for face super resolution problem, such as GFPGAN \cite{GFPGAN}, GPEN \cite{GPEN}, HiFaceGAN \cite{HiFaceGAN}, DFDNet \cite{DFDNet}, PSFRGAN \cite{PSFRGAN}, GCFSR \cite{GCFSR}. Among these state-of-the-art models, the GCFSR \cite{GCFSR} has achieved higher performance in the terms of PSNR and DFDNet has acchieved higher performance in the terms of SSIM on CelebA-HQ test dataset for 4x super-resolution. The quantitative evaluation for FSR problem for 4x is shown in Table \ref{tab:PSNR/SSIM_FSR}. 
Our results outperform the state-of-the-art models in both qualitative and quantitative aspects, as demonstrated by higher SSIM and PSNR scores. Furthermore, our model achieves comparable results on the FID metric. This suggests that our model is not only better at reconstructing high-resolution images but also produces images that are more faithful to the original.

\subsection{Qualitative Evaluation} 
For the qualitative evaluation we compare the generated images using SRTransGAN method with the generated images using existing methods to observe the visual quality. The image super-resolution results for different models such as HR (Ground Truth), CARN \cite{CARN}, ESRT \cite{ESRT}, SAN \cite{SAN}, IGNN \cite{IGNN}, IPT \cite{IPT}, RCAN \cite{RCAN} and the proposed SRTransGAN over Set5, Set14, BSD100, and Urban100 datasets are shown in Fig. \ref{fig:datasetx2} for $2\times$ super-resolution and Fig. \ref{fig:datasetx4} for $4\times$ super-resolution.
In Fig. \ref{fig:datasetx2} four images are taken as example form different datasets such as `Bird' image from Set5, `Babloon' image from Set14 and `Img\_014' image from Urban100 dataset. Here, first column shows the `LR' images and rest columns from left to right show the generated `SR' (super-resolution) images using HR, CARN, ESRT, SAN models in first row and IGNN, IPT, RCAN, SRTransGAN models in second row, respectively. The better visual quality can be observed in the SR images generated by SRTransGAN model at $2\times$ scale.
In Fig. \ref{fig:datasetx4} four images are taken as example form different datasets such as `Butterfly' image from Set5, `Flowers' image from Set14 and `Img\_012' image from Urban100 dataset. Here, first column shows the `LR' images and rest columns from left to right show the generated `SR' (super-resolution) images using HR, CARN, ESRT, SAN models in first row and IGNN, IPT, RCAN, SRTransGAN models in second row, respectively. We also observe better SR images by the proposed model at $4\times$ scale.
The proposed SRTransGAN model performs well in terms of qualitative evaluation (i.e., visual quality) as we utilize the transformer encoder-decoder based generator network in multi-level fashion dealing the scales of visual appearance effectively. Moreover, the use adversarial loss with conditional inputs also play a vital role to generate the visually appealing super-resolution images. 



\begin{figure}[!t]
\centering
\includegraphics[trim={1cm 5mm 2mm 0},clip,width=.9\columnwidth]{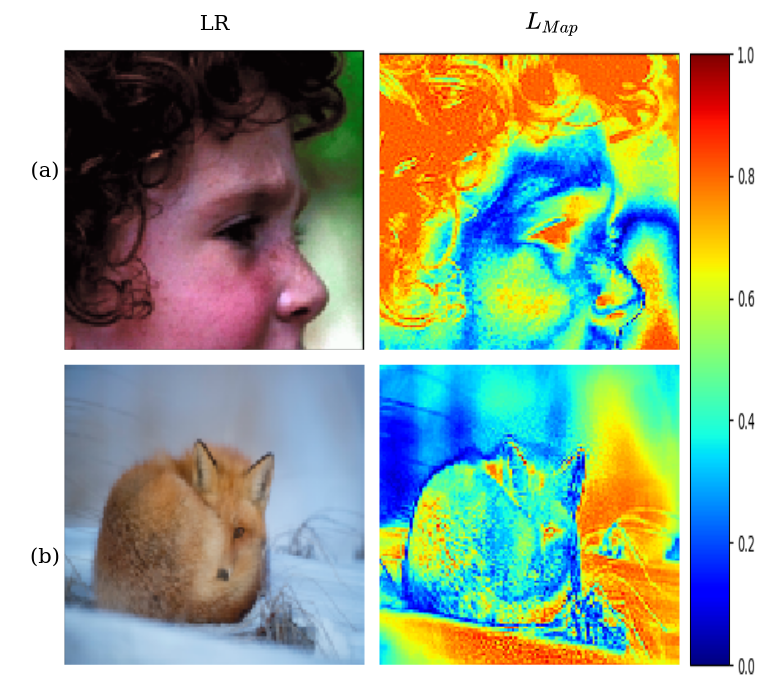}
\caption{Visual Activation Maps for Proposed SRTransGAN (Multi-Color view for best Visual representation). }
\label{fig:saliency}
\end{figure}


\subsection{Visual Activation Map}
The visual activation map ($L_{Map}$) highlights the important regions in input images. It is calculated as the gradient by back-propagating the HR image through the generator network ($SRTransG$) while the corresponding LR image is used as input to generator during forward pass. Finally, the generated gradient map is normalized in between 0 and 1 and used as the visual activation map. 
The $L_{Map} \in \mathbb{R}^{H,W,1}$ is expressed as \cite{vanilagradsaliency}:
\begin{equation}
L_{Map} = \eta ( \nabla (SRTransG(LR)) )\sim (0,1)
\end{equation}
where, $\nabla$ is gradient at LR image which is the cost difference between HR and SR images and $\eta$ is normalisation in $(0,1)$.

The $L_{Map}$ of LR image is shown in Fig. \ref{fig:saliency}. The $L_{Map}$ is represented in multi-color for the better visualization. The blue color represents the regions where model gets lower gradients (i.e., least important) and the red color represents the regions where model gets higher gradients (i.e., most important) in terms of the image super-resolution. It is evident that the high frequency regions are very important for the image super-resolution problem. 
In first example of Fig. \ref{fig:saliency} it can be observed that the model focuses more on the person's hair because it contains hair-lined structure (i.e., edges) which is a high frequency information. The model also focuses on eyes and near nose regions which can be distorted due to blur in the LR image.  
A similar observation is also perceived in the second example where model focuses on the right background blurred regions as well as on the eye of animal because it contains minute edges over the animal regions. 
It can be observed that the proposed method learns more at the edges and blur regions while is essential for any effective image super-resolution method.

\section{Conclusion} \label{sec:Conclusion}
In this work, we propose a novel image super-resolution Transformer GAN (SRTransGAN) architecture. The proposed generator network utilizes the transformer encoder, decoder and refinement modules with the encoder and decoder modules forming a multilevel feature learning with cross-scale information in feature space in a progressive manner. The discriminator network exploits the discriminative ability of vision transformer and provides a strong adversarial competence for the better training of generator network. It is noticed from the visualization of gradients that the proposed model is able to super scale the edges and blur regions very effectively leading to visually appealing super-resolution output images. It is also observed experimentally that the proposed SRTransGAN outperforms the state-of-the-art models with a great margin on most of the benchmark single image super-resolution datasets for both $2\times$ as well as $4\times$ super-resolution. The SRTransGAN model leads to an average \% improvement by 7.12\% and 2.14\% in terms of PSNR and SSIM, respectively, for $2\times$ SR task and  by 4.76\% and 3.51\%, respectively, for $4\times$ SR task, as compared to the best competitive models. The higher levels in the proposed encoder-decoder architecture of  transformer based generator lead to better performance due to the better encoding of features at different and cross-scales. Ablation study suggests that the performance of the proposed model can be further improved by increasing the size and complexity of the training dataset. The proposed model can be easily extended to different type of GAN architectures and different image processing applications.

 



\bibliographystyle{IEEEtran}
\bibliography{main}

\begin{thebibliography}{10}
\providecommand{\url}[1]{#1}
\csname url@samestyle\endcsname
\providecommand{\newblock}{\relax}
\providecommand{\bibinfo}[2]{#2}
\providecommand{\BIBentrySTDinterwordspacing}{\spaceskip=0pt\relax}
\providecommand{\BIBentryALTinterwordstretchfactor}{4}
\providecommand{\BIBentryALTinterwordspacing}{\spaceskip=\fontdimen2\font plus
\BIBentryALTinterwordstretchfactor\fontdimen3\font minus \fontdimen4\font\relax}
\providecommand{\BIBforeignlanguage}[2]{{%
\expandafter\ifx\csname l@#1\endcsname\relax
\typeout{** WARNING: IEEEtran.bst: No hyphenation pattern has been}%
\typeout{** loaded for the language `#1'. Using the pattern for}%
\typeout{** the default language instead.}%
\else
\language=\csname l@#1\endcsname
\fi
#2}}
\providecommand{\BIBdecl}{\relax}
\BIBdecl

\bibitem{ed_chen2021multispectral}
W.~Chen, Z.~Jia, J.~Yang, and N.~K. Kasabov, ``Multispectral image enhancement based on weighted principal component analysis and improved fractional differential mask,'' \emph{IEEE Geoscience and Remote Sensing Letters}, vol.~19, pp. 1--5, 2021.

\bibitem{ed_huang2021effective}
Z.~Huang, Z.~Jia, J.~Yang, and N.~K. Kasabov, ``An effective algorithm for specular reflection image enhancement,'' \emph{IEEE Access}, vol.~9, pp. 154\,513--154\,523, 2021.

\bibitem{survey2020IeeeTransPatten}
Z.~Wang, J.~Chen, and S.~C. Hoi, ``Deep learning for image super-resolution: A survey,'' \emph{IEEE transactions on pattern analysis and machine intelligence}, vol.~43, no.~10, pp. 3365--3387, 2020.

\bibitem{ieee_imsur}
Q.~Jiang, Z.~Liu, K.~Gu, F.~Shao, X.~Zhang, H.~Liu, and W.~Lin, ``Single image super-resolution quality assessment: a real-world dataset, subjective studies, and an objective metric,'' \emph{IEEE Transactions on Image Processing}, vol.~31, pp. 2279--2294, 2022.

\bibitem{SRCNNr1}
C.~Dong, C.~C. Loy, K.~He, and X.~Tang, ``Image super-resolution using deep convolutional networks,'' \emph{IEEE transactions on pattern analysis and machine intelligence}, vol.~38, no.~2, pp. 295--307, 2015.

\bibitem{FSRCNNr2}
C.~Dong, C.~C. Loy, and X.~Tang, ``Accelerating the super-resolution convolutional neural network,'' in \emph{European conference on computer vision}.\hskip 1em plus 0.5em minus 0.4em\relax Springer, 2016, pp. 391--407.

\bibitem{VDSRr3}
J.~Kim, J.~K. Lee, and K.~M. Lee, ``Accurate image super-resolution using very deep convolutional networks,'' in \emph{Proceedings of the IEEE conference on computer vision and pattern recognition}, 2016, pp. 1646--1654.

\bibitem{Memnet}
Y.~Tai, J.~Yang, X.~Liu, and C.~Xu, ``Memnet: A persistent memory network for image restoration,'' in \emph{Proceedings of the IEEE international conference on computer vision}, 2017, pp. 4539--4547.

\bibitem{SRGAN}
C.~Ledig, L.~Theis, F.~Husz{\'a}r, J.~Caballero, A.~Cunningham, A.~Acosta, A.~Aitken, A.~Tejani, J.~Totz, Z.~Wang \emph{et~al.}, ``Photo-realistic single image super-resolution using a generative adversarial network,'' in \emph{Proceedings of the IEEE conference on computer vision and pattern recognition}, 2017, pp. 4681--4690.

\bibitem{esrgan}
X.~Wang, K.~Yu, S.~Wu, J.~Gu, Y.~Liu, C.~Dong, Y.~Qiao, and C.~Change~Loy, ``Esrgan: Enhanced super-resolution generative adversarial networks,'' in \emph{Proceedings of the European conference on computer vision (ECCV) workshops}, 2018, pp. 0--0.

\bibitem{GMGAN}
X.~Zhu, L.~Zhang, L.~Zhang, X.~Liu, Y.~Shen, and S.~Zhao, ``Gan-based image super-resolution with a novel quality loss,'' \emph{Mathematical Problems in Engineering}, 2020.

\bibitem{ESRT}
Z.~Lu, H.~Liu, J.~Li, and L.~Zhang, ``Efficient transformer for single image super-resolution,'' in \emph{Proceedings of the IEEE/CVF Conference on Computer Vision and Pattern Recognition Workshop}, 2022.

\bibitem{RCAN}
Y.~Zhang, K.~Li, K.~Li, L.~Wang, B.~Zhong, and Y.~Fu, ``Image super-resolution using very deep residual channel attention networks,'' in \emph{Proceedings of the European conference on computer vision (ECCV)}, 2018, pp. 286--301.

\bibitem{IPT}
H.~Chen, Y.~Wang, T.~Guo, C.~Xu, Y.~Deng, Z.~Liu, S.~Ma, C.~Xu, C.~Xu, and W.~Gao, ``Pre-trained image processing transformer,'' in \emph{Proceedings of the IEEE/CVF Conference on Computer Vision and Pattern Recognition}, 2021, pp. 12\,299--12\,310.

\bibitem{GCFSR}
J.~He, W.~Shi, K.~Chen, L.~Fu, and C.~Dong, ``Gcfsr: a generative and controllable face super resolution method without facial and gan priors,'' in \emph{Proceedings of the IEEE/CVF Conference on Computer Vision and Pattern Recognition}, 2022, pp. 1889--1898.

\bibitem{SpatiospectralSR}
Q.~Ma, J.~Jiang, X.~Liu, and J.~Ma, ``Multi-task interaction learning for spatiospectral image super-resolution,'' \emph{IEEE Transactions on Image Processing}, 2022.

\bibitem{ThermalSR}
H.~Gupta and K.~Mitra, ``Toward unaligned guided thermal super-resolution,'' \emph{IEEE Transactions on Image Processing}, vol.~31, pp. 433--445, 2021.

\bibitem{EDSR-baseline}
B.~Lim, S.~Son, H.~Kim, S.~Nah, and K.~Mu~Lee, ``Enhanced deep residual networks for single image super-resolution,'' in \emph{Proceedings of the IEEE conference on computer vision and pattern recognition workshops}, 2017, pp. 136--144.

\bibitem{RDN}
Y.~Zhang, Y.~Tian, Y.~Kong, B.~Zhong, and Y.~Fu, ``Residual dense network for image super-resolution,'' in \emph{Proceedings of the IEEE conference on computer vision and pattern recognition}, 2018, pp. 2472--2481.

\bibitem{DRRN}
Y.~Tai, J.~Yang, and X.~Liu, ``Image super-resolution via deep recursive residual network,'' in \emph{Proceedings of the IEEE conference on computer vision and pattern recognition}, 2017, pp. 3147--3155.

\bibitem{tdpn}
Q.~Cai, J.~Li, H.~Li, Y.-H. Yang, F.~Wu, and D.~Zhang, ``Tdpn: Texture and detail-preserving network for single image super-resolution,'' \emph{IEEE Transactions on Image Processing}, vol.~31, pp. 2375--2389, 2022.

\bibitem{WBTRN}
Z.~Li, Z.-S. Kuang, Z.-L. Zhu, H.-P. Wang, and X.-L. Shao, ``Wavelet-based texture reformation network for image super-resolution,'' \emph{IEEE Transactions on Image Processing}, vol.~31, pp. 2647--2660, 2022.

\bibitem{MPRN}
Q.~Wang, Q.~Gao, L.~Wu, G.~Sun, and L.~Jiao, ``Adversarial multi-path residual network for image super-resolution,'' \emph{IEEE Transactions on Image Processing}, vol.~30, pp. 6648--6658, 2021.

\bibitem{CARN}
N.~Ahn, B.~Kang, and K.-A. Sohn, ``Fast, accurate, and lightweight super-resolution with cascading residual network,'' in \emph{Proceedings of the European Conference on Computer Vision (ECCV)}, 2018, pp. 252--268.

\bibitem{IMDN}
Z.~Hui, X.~Gao, Y.~Yang, and X.~Wang, ``Lightweight image super-resolution with information multi-distillation network,'' in \emph{Proceedings of the 27th ACM International Conference on Multimedia}, 2019, pp. 2024--2032.

\bibitem{LapSRN}
W.-S. Lai, J.-B. Huang, N.~Ahuja, and M.-H. Yang, ``Fast and accurate image super-resolution with deep laplacian pyramid networks,'' \emph{IEEE transactions on pattern analysis and machine intelligence}, vol.~41, no.~11, pp. 2599--2613, 2018.

\bibitem{DRCN}
J.~Kim, J.~K. Lee, and K.~M. Lee, ``Deeply-recursive convolutional network for image super-resolution,'' in \emph{Proceedings of the IEEE conference on computer vision and pattern recognition}, 2016, pp. 1637--1645.

\bibitem{SRMDNF}
K.~Zhang, W.~Zuo, and L.~Zhang, ``Learning a single convolutional super-resolution network for multiple degradations,'' in \emph{Proceedings of the IEEE Conference on Computer Vision and Pattern Recognition}, 2018, pp. 3262--3271.

\bibitem{RNAN}
Y.~Zhang, K.~Li, K.~Li, B.~Zhong, and Y.~Fu, ``Residual non-local attention networks for image restoration,'' \emph{arXiv preprint arXiv:1903.10082}, 2019.

\bibitem{OISR-RK3}
X.~He, Z.~Mo, P.~Wang, Y.~Liu, M.~Yang, and J.~Cheng, ``Ode-inspired network design for single image super-resolution,'' in \emph{Proceedings of the IEEE/CVF Conference on Computer Vision and Pattern Recognition}, 2019, pp. 1732--1741.

\bibitem{robust_tvtv}
M.~Vella and J.~F. Mota, ``Robust single-image super-resolution via cnns and tv-tv minimization,'' \emph{IEEE Transactions on Image Processing}, vol.~30, pp. 7830--7841, 2021.

\bibitem{srgat}
Y.~Yan, W.~Ren, X.~Hu, K.~Li, H.~Shen, and X.~Cao, ``Srgat: Single image super-resolution with graph attention network,'' \emph{IEEE Transactions on Image Processing}, vol.~30, pp. 4905--4918, 2021.

\bibitem{IGNN}
S.~Zhou, J.~Zhang, W.~Zuo, and C.~C. Loy, ``Cross-scale internal graph neural network for image super-resolution,'' \emph{Advances in neural information processing systems}, vol.~33, pp. 3499--3509, 2020.

\bibitem{PIX}
P.~Isola, J.-Y. Zhu, T.~Zhou, and A.~A. Efros, ``Image-to-image translation with conditional adversarial networks,'' in \emph{Proceedings of the IEEE conference on computer vision and pattern recognition}, 2017, pp. 1125--1134.

\bibitem{pcsgan}
K.~K. Babu and S.~R. Dubey, ``Pcsgan: perceptual cyclic-synthesized generative adversarial networks for thermal and nir to visible image transformation,'' \emph{Neurocomputing}, vol. 413, pp. 41--50, 2020.

\bibitem{DUSGAN}
K.~Prajapati, V.~Chudasama, H.~Patel, K.~Upla, K.~Raja, R.~Ramachandra, and C.~Busch, ``Direct unsupervised super-resolution using generative adversarial network (dus-gan) for real-world data,'' \emph{IEEE Transactions on Image Processing}, vol.~30, pp. 8251--8264, 2021.

\bibitem{attention2017vaswani}
A.~Vaswani, N.~Shazeer, N.~Parmar, J.~Uszkoreit, L.~Jones, A.~N. Gomez, {\L}.~Kaiser, and I.~Polosukhin, ``Attention is all you need,'' \emph{Advances in neural information processing systems}, vol.~30, 2017.

\bibitem{khan2021transformers}
S.~Khan, M.~Naseer, M.~Hayat, S.~W. Zamir, F.~S. Khan, and M.~Shah, ``Transformers in vision: A survey,'' \emph{ACM Computing Surveys (CSUR)}, 2021.

\bibitem{vit}
A.~Dosovitskiy, L.~Beyer, A.~Kolesnikov, D.~Weissenborn, X.~Zhai, T.~Unterthiner, M.~Dehghani, M.~Minderer, G.~Heigold, S.~Gelly \emph{et~al.}, ``An image is worth 16x16 words: Transformers for image recognition at scale,'' \emph{arXiv preprint arXiv:2010.11929}, 2020.

\bibitem{fan2021multiscale}
H.~Fan, B.~Xiong, K.~Mangalam, Y.~Li, Z.~Yan, J.~Malik, and C.~Feichtenhofer, ``Multiscale vision transformers,'' in \emph{Proceedings of the IEEE/CVF International Conference on Computer Vision}, 2021, pp. 6824--6835.

\bibitem{SAN}
T.~Dai, J.~Cai, Y.~Zhang, S.-T. Xia, and L.~Zhang, ``Second-order attention network for single image super-resolution,'' in \emph{Proceedings of the IEEE/CVF conference on computer vision and pattern recognition}, 2019, pp. 11\,065--11\,074.

\bibitem{vitgan}
K.~Lee, H.~Chang, L.~Jiang, H.~Zhang, Z.~Tu, and C.~Liu, ``Vitgan: Training gans with vision transformers,'' \emph{arXiv preprint arXiv:2107.04589}, 2021.

\bibitem{Transgan}
Y.~Jiang, S.~Chang, and Z.~Wang, ``Transgan: Two pure transformers can make one strong gan, and that can scale up,'' \emph{Advances in Neural Information Processing Systems}, vol.~34, 2021.

\bibitem{restormer2021zamir}
S.~W. Zamir, A.~Arora, S.~Khan, M.~Hayat, F.~S. Khan, and M.-H. Yang, ``Restormer: Efficient transformer for high-resolution image restoration,'' in \emph{Proceedings of the IEEE conference on computer vision and pattern recognition}, 2022.

\bibitem{gelu}
D.~Hendrycks and K.~Gimpel, ``Gaussian error linear units (gelus),'' \emph{arXiv preprint arXiv:1606.08415}, 2016.

\bibitem{layerNORM}
J.~L. Ba, J.~R. Kiros, and G.~E. Hinton, ``Layer normalization,'' \emph{arXiv preprint arXiv:1607.06450}, 2016.

\bibitem{Dropout}
N.~Srivastava, G.~Hinton, A.~Krizhevsky, I.~Sutskever, and R.~Salakhutdinov, ``Dropout: a simple way to prevent neural networks from overfitting,'' \emph{The journal of machine learning research}, vol.~15, no.~1, pp. 1929--1958, 2014.

\bibitem{GFPGAN}
X.~Wang, Y.~Li, H.~Zhang, and Y.~Shan, ``Towards real-world blind face restoration with generative facial prior,'' in \emph{Proceedings of the IEEE/CVF Conference on Computer Vision and Pattern Recognition}, 2021, pp. 9168--9178.

\bibitem{GPEN}
T.~Yang, P.~Ren, X.~Xie, and L.~Zhang, ``Gan prior embedded network for blind face restoration in the wild,'' in \emph{Proceedings of the IEEE/CVF Conference on Computer Vision and Pattern Recognition}, 2021, pp. 672--681.

\bibitem{HiFaceGAN}
L.~Yang, S.~Wang, S.~Ma, W.~Gao, C.~Liu, P.~Wang, and P.~Ren, ``Hifacegan: Face renovation via collaborative suppression and replenishment,'' in \emph{Proceedings of the 28th ACM international conference on multimedia}, 2020, pp. 1551--1560.

\bibitem{DFDNet}
X.~Li, C.~Chen, S.~Zhou, X.~Lin, W.~Zuo, and L.~Zhang, ``Blind face restoration via deep multi-scale component dictionaries,'' in \emph{Computer Vision--ECCV 2020: 16th European Conference, Glasgow, UK, August 23--28, 2020, Proceedings, Part IX 16}.\hskip 1em plus 0.5em minus 0.4em\relax Springer, 2020, pp. 399--415.

\bibitem{PSFRGAN}
C.~Chen, X.~Li, L.~Yang, X.~Lin, L.~Zhang, and K.-Y.~K. Wong, ``Progressive semantic-aware style transformation for blind face restoration,'' in \emph{Proceedings of the IEEE/CVF conference on computer vision and pattern recognition}, 2021, pp. 11\,896--11\,905.

\bibitem{datadiv}
R.~Timofte, E.~Agustsson, L.~Van~Gool, M.-H. Yang, and L.~Zhang, ``Ntire 2017 challenge on single image super-resolution: Methods and results,'' in \emph{Proceedings of the IEEE conference on computer vision and pattern recognition workshops}, 2017, pp. 114--125.

\bibitem{set5}
M.~Bevilacqua, A.~Roumy, C.~Guillemot, and M.~L. Alberi-Morel, ``Low-complexity single-image super-resolution based on nonnegative neighbor embedding,'' in \emph{Proceedings of the 23rd British Machine Vision Conference (BMVC)}.\hskip 1em plus 0.5em minus 0.4em\relax BMVA press, 2012.

\bibitem{set14}
J.~Yang, J.~Wright, T.~S. Huang, and Y.~Ma, ``Image super-resolution via sparse representation,'' \emph{IEEE Transactions on image processing}, vol.~19, no.~11, pp. 2861--2873, 2010.

\bibitem{bsd100}
D.~Martin, C.~Fowlkes, D.~Tal, and J.~Malik, ``A database of human segmented natural images and its application to evaluating segmentation algorithms and measuring ecological statistics,'' in \emph{Proceedings Eighth IEEE International Conference on Computer Vision. ICCV 2001}, vol.~2.\hskip 1em plus 0.5em minus 0.4em\relax IEEE, 2001, pp. 416--423.

\bibitem{urban100}
J.-B. Huang, A.~Singh, and N.~Ahuja, ``Single image super-resolution from transformed self-exemplars,'' in \emph{Proceedings of the IEEE conference on computer vision and pattern recognition}, 2015, pp. 5197--5206.

\bibitem{datacufed}
Z.~Zhang, Z.~Wang, Z.~Lin, and H.~Qi, ``Image super-resolution by neural texture transfer,'' in \emph{Proceedings of the IEEE/CVF Conference on Computer Vision and Pattern Recognition}, 2019, pp. 7982--7991.

\bibitem{goodfellow2014generative}
I.~Goodfellow, J.~Pouget-Abadie, M.~Mirza, B.~Xu, D.~Warde-Farley, S.~Ozair, A.~Courville, and Y.~Bengio, ``Generative adversarial nets,'' \emph{Advances in neural information processing systems}, vol.~27, 2014.

\bibitem{vanilagradsaliency}
K.~Simonyan, A.~Vedaldi, and A.~Zisserman, ``Deep inside convolutional networks: Visualising image classification models and saliency maps,'' in \emph{In Workshop at International Conference on Learning Representations}.\hskip 1em plus 0.5em minus 0.4em\relax Citeseer, 2014.

\end{thebibliography}
\end{document}